\documentclass[final,12pt]{aipproc}
\layoutstyle{6x9}

\usepackage{graphicx}% Include figure files
\usepackage{bm}% bold math
\usepackage[centertags]{amsmath}
\usepackage{amssymb}
\usepackage{longtable}

\def\mbsix{\mathbf{6}}
\def\mbsixb{\mathbf{\overline{6}}}
\def\mbthree{\mathbf{3}}
\def\mbthreeb{\mathbf{\overline{3}}}

\begin{document}

\title{String Model Building}


\author{Stuart Raby}{
address={Physics Department, The Ohio State University, 191 W.
Woodruff Ave.,Columbus, OH 43210, USA} }

\begin{abstract}
In this talk I review some recent progress in heterotic and F theory
model building. I then consider work in progress attempting to find
the F theory dual to a class of heterotic orbifold models which come
quite close to the MSSM.
\end{abstract}

\maketitle

%%%%%%%%%%%%%%%%%%%%%%%%%%%%%%%%%%%%%%%%%%%%
%% MAINMATTER
%%%%%%%%%%%%%%%%%%%%%%%%%%%%%%%%%%%%%%%%%%%%

\section{Random searches for the MSSM}
The literature includes a record of several searches for the MSSM.
Among these have been random searches in the string landscape
looking for features common to the MSSM.  In particular, vacua with
N = 1 supersymmetry [SUSY],  the Standard Model gauge group and
three families of quarks and leptons.   These random searches have
for the most part shown that the MSSM is an extremely rare point in
the string landscape.   For example, searches in Type II
intersecting D-brane models~\cite{Gmeiner:2005vz} have found nothing
looking like the MSSM in $10^9$ tries.  Searches in Gepner
orientifolds have been a bit more successful finding one MSSM-like
model for every 10,000 tries~\cite{Anastasopoulos:2006da}. Even
searches in the heterotic string, using the free fermionic
construction,  have shown that the MSSM is a very rare point in the
string landscape~\cite{Dienes:2007ms}.  The bottom line: if you want
to find the MSSM, then a random search is not the way to go.   In
fact, MSSM-like models have been found in Type II D-brane vacua, BUT
by directed searches~\cite{Blumenhagen:2005mu} AND not random ones.

\subsection{Virtues of SUSY GUTs}

We will propose a very particular directed search.   We will require
that SUSY GUTs be incorporated at the first step.  This constraint
is motivated by the many virtues of SUSY GUTs.
\begin{enumerate}
\item They can ``naturally" explain why the weak scale $M_Z << M_{GUT}$;
\item Explains  charge  quantization;
\item Predicts gauge coupling unification  !!
\item Predicts SUSY particles at LHC !
\item Predicts proton decay;
\item Predicts Yukawa coupling unification,
\item and with Family symmetry can explain the Fermion mass hierarchy;
\item Explains neutrino masses and mixing via See-Saw;
\item The LSP is a Dark Matter candidate, and
\item Can give a cosmological asymmetry in the number of baryons minus
anti-baryons, i.e. baryogenesis via leptogenesis.
\end{enumerate}
For all of these reasons we might suspect that SUSY GUTs are a
fundamental component of any realistic string vacuum.   And thus if
one is searching for the MSSM in the mostly barren string landscape,
one should incorporate SUSY GUTs at the first step.

\section{Heterotic model building}

I will not attempt to discuss the many attempts to find the MSSM in
the string landscape.  Instead let me just discuss some recent
progress in heterotic model building, either on a smooth Calabi-Yau
3-fold or in the context of orbifold contructions.

\subsection{Smooth Manifolds}

\begin{figure}
  \includegraphics[height=3cm]{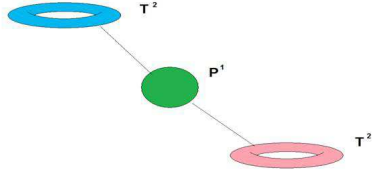}
  \caption{Calabi-Yau 3-fold defined in terms of a double elliptic fibration. The line connecting
  the tori to the 2-sphere represents the fibration.}
  \label{fig:CY3}
\end{figure}

Bouchard et al.~\cite{Bouchard:2005ag}  have obtained an $SU(5)$ GUT
model on a CY$_3$ with the following properties.  They have three
families of quarks and leptons, and one or two pairs of Higgs
doublets. They accomplish GUT symmetry breaking and Higgs
doublet-triplet splitting via a Wilson line in the weak hypercharge
direction.  The CY$_3$ is defined by a double elliptic fibration,
i.e. two tori whose radii change as the tori move over the surface
of a sphere (see Fig. \ref{fig:CY3}).

In addition, they obtain a non-trivial up Yukawa matrix given
by~\cite{Bouchard:2006dn}
\begin{equation}
\lambda_u = \left( \begin{array}{ccc} a & b & c \\ b & d & e \\ c & e & 0
\end{array} \right) .
\end{equation}
The parameters $a, \cdots , e$ are functions of the moduli.
The down and charged lepton Yukawa matrices are however zero and
would require non-perturbative effects to change this.

\subsection{Orbifolds}

Early work on orbifold constructions of the heterotic string was
started over 20 years
ago~\cite{Dixon:1985jw,Dixon:1986jc,Ibanez:1986tp,Ibanez:1987sn,Font:1989aj}.
However, progress has been made
recently~\cite{Kobayashi:2004ud,Kobayashi:2004ya,Buchmuller:2005jr,Buchmuller:2006ik,Lebedev:2006kn,Lebedev:2006tr,Lebedev:2007hv,Kim:2006hv,Kim:2006zw,Kim:2007mt}.
In a ``mini-landscape" search of the $E(8) \times E(8)$ heterotic
landscape~\cite{Lebedev:2007hv} 223 models with 3 families, Higgs
doublets and ONLY vector-like exotics were found out of a total of
order 30,000 models or approximately 1 in 100 models searched looked
like the MSSM!  We called this a ``fertile patch" in the heterotic
landscape. Let me describe this focussed search in more detail.

\begin{figure}
  \includegraphics[height=2cm]{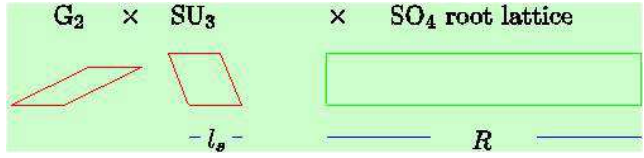}
  \caption{Product of 3 two dimensional tori defined in terms of the root lattice of the
  respective Lie groups.}
  \label{fig:tori}
\end{figure}

We compactify the $E(8) \times E(8)$ heterotic string on the product
of 3 two dimensional tori defined in terms of the root lattice of a
$G(2)$, $SU(3)$ and $SO(4)$ root lattices (see Fig. \ref{fig:tori}).
These tori are chosen since they are invariant under the symmetry
${\mathbb Z}_6-II = {\mathbb Z}_3 \times {\mathbb Z}_2$ which
defines the orbifold.  We also imbed the orbifold symmetry into the
$E(8) \times E(8)$ gauge lattice via a shift vector, $V_6$ and add
Wilson lines, consistent with modular invariance. This has the
effect of breaking the gauge group to a subgroup (without breaking
the rank).   Note 5 cycles on the tori are small of order the string
length, while one cycle in the $SO(4)$ torus is assumed to be large.

\begin{figure}
  \includegraphics[height=1.2cm]{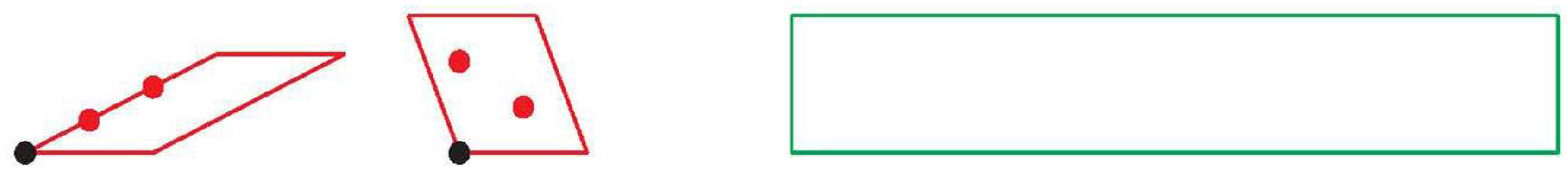}
  \caption{Product of 3 two dimensional tori with ${\mathbb Z}_3$ orbifold fixed points explicit.}
  \label{fig:torimodz3}
\end{figure}

First consider the action of the ${\mathbb Z}_3$ orbifold with a
Wilson line in the $SU(3)$ torus (see Fig.\ref{fig:torimodz3}).  Let
us focus on the states transforming under the visible sector $E(8)$.
The massless states from the untwisted sector and $(G(2), SU(3))$
twisted sectors transform as N=1 super-multiplets in 6 dimensions
(or N=2 in terms of effective 4 dimensional super-multiplets) as a
vector hyper-multiplet in the adjoint representation of $SU(6)$ and
chiral hyper-multiplets in the $20 +  9(6 + \bar 6)$ dimensional
representations.   All of these states move freely in the untwisted
torus.  Thus they describe the degrees of freedom for a 6
dimensional $SU(6)$ orbifold GUT.

\begin{figure}
  \includegraphics[height=3.5cm]{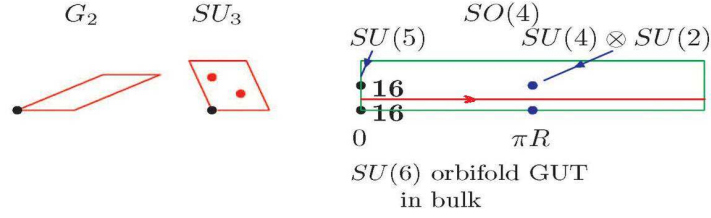}
  \caption{Product of 3 two dimensional tori with ${\mathbb Z}_6$ orbifold fixed points explicit.}
  \label{fig:torimodz6}
\end{figure}

\begin{figure}
  \includegraphics[height=5.5cm]{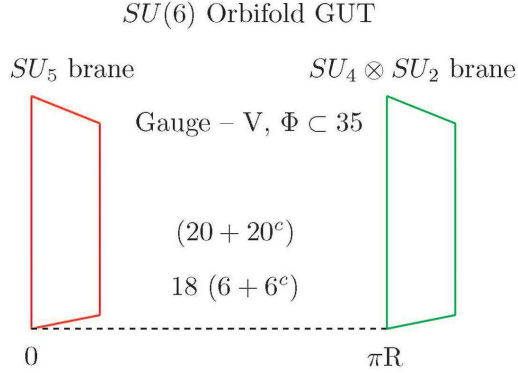}
  \caption{Viewed in terms of 5 dimensional orbifold GUT field theory with and $SU(5)$ brane on the left
  and an $SU(4) \times SU(2)$  brane on the right.  The overlap is the Standard Model gauge group.}
  \label{fig:orbifold}
\end{figure}

We then perform the additional ${\mathbb Z}_2$ orbifold and add a
Wilson line along the long cycle in the $SO(4)$ torus (see Fig.
\ref{fig:torimodz6}). The resulting theory has the Standard Model
gauge symmetry in the visible sector.   The shift vector $V_2$ acts
as a parity P on the left, breaking $SU(6)$ to $SU(5)$.   On the
right, the combination of $V_2 + W_2$ gives a parity P$^\prime$
which breaks further to the Standard Model gauge group.   In
addition, two light families are located at the two $SO(4)$ fixed
points. In Fig. \ref{fig:orbifold} this is represented in terms of
an effective 5 dimensional orbifold $SU(6)$ GUT field theory.
Note, the two $SO(4)$ fixed points are very special.   In terms of
the full 10D string theory,  these correspond to ``local" $SO(10)$
fixed points.  Thus families come in complete $SO(10)$
representations.   This is NOT an accident, but was enforced from
the very beginning.   The Higgs and third family come from the  5D
``bulk."   They can be seen as follows.   First consider the vector
hypermultiplet.   In terms of 6 x 6 hermitian matrices we have the
gauge multiplet on the left (Eqn. \ref{eq:vector}) and the chiral
adjoint on the right.  The subscripts label the charge under the
parity (P, P$^\prime$).  All states include Kaluza-Klein modes
beginning at the compactification scale $M_C = 1/R$, while only
$(++)$ states contain massless modes.   On the left, these include
the $SU(3) \times SU(2) \times U(1)_Y$ SM  gauge sector.   On the
right we have one pair of Higgs doublets.   Note,  the orbifolding
has succeeded in splitting the Higgs doublets and triplets.  This
model is a string theory realization of gauge-Higgs unification.
\begin{equation}  \label{eq:vector}
\left( \begin{array}{ccc} V_{++}^{3 \times 3} & V_{+-}^{3 \times 2} & V_{-+}^{3 \times 1} \\
V_{+-}^{2 \times 3} & V_{++}^{2 \times 2} & V_{--}^{2 \times 1} \\
V_{-+}^{1 \times 3} & V_{--}^{1 \times 2} & V_{++}^{1 \times 1} \end{array} \right) ,
\left( \begin{array}{ccc} \Phi_{--}^{3 \times 3} & \Phi_{-+}^{3 \times 2} & \Phi_{+-}^{3 \times 1} \\
\Phi_{-+}^{2 \times 3} & \Phi_{--}^{2 \times 2} & \Phi_{++}^{2 \times 1} [= H_u] \\
\Phi_{+-}^{1 \times 3} & \Phi_{++}^{1 \times 2} [= H_d] & \Phi_{--}^{1 \times 1} \end{array} \right)
\end{equation}
In addition the third family is contained in [in terms of effective
4D N=1 chiral superfields]
\begin{eqnarray}  ( 20 + 20^c ) & \supset & Q_3 + \bar t + \bar \tau \\
2 ( 6 + 6^c ) & \supset & L_3 + \bar b  \nonumber
\end{eqnarray}
As a result we obtain gauge-Yukawa coupling unification with
\begin{equation}
W \supset  \frac{g_5}{\sqrt{\pi R}} \int_0^{\pi R} dy 20^c \Phi 20 = g_{GUT} Q_3 H_u \bar t
\end{equation}

The model also has a discrete non-Abelian family symmetry,  $D_4$.
The symmetry acts on the two light families as doublets and the
third family and Higgs doublets are singlets under $D_4$.   The
order 8 group, $D_4$, is generated by the two operations given by $
\sigma_1 = \left( \begin{array}{cc} 0 & 1 \\ 1 & 0 \end{array}
\right)$  and $\sigma_3 = \left( \begin{array}{cc} 1 & 0 \\ 0 & -1
\end{array} \right)$.   The first is a symmetry which interchanges
the two light families located at opposite sides of one cycle of the
orbifolded $SO(4)$ torus.   This symmetry corresponds to the
geometric translation half way around the cycle which is not broken
by the Wilson line lying along the orthogonal cycle.   The second is
a result of so-called space-group selection rules which require an
even number of states at each of these two fixed points. As a
result, the theory is invariant under the action of multiplying each
state located at, say, the lower fixed point by minus one.  As a
consequence of $D_4$ (and additional U(1) symmetries) only the third
family can obtain a tree level Yukawa coupling.  All others Yukawa
couplings can only be obtained once the family symmetries are
broken.  Thus the string theory includes a natural Froggatt-Nielsen
mechanism~\cite{Froggatt:1978nt} for generating a hierarchy of
fermion masses.   [Aside, for a general analysis of discrete
non-Abelian flavor symmetries obtainable in orbifold models,
see~\cite{Kobayashi:2006wq} and for the phenomenological analysis of
a $D_4$ invariant description of quark masses and flavor violation,
see~\cite{Ko:2007dz}.]

\begin{table}[h]
\centerline{%
\begin{tabular}{|c|l|l|c|c|l|l|}
\hline
\# & irrep & label & & \# & irrep & label\\
\hline
 3 &
$\left(\boldsymbol{3},\boldsymbol{2};\boldsymbol{1},\boldsymbol{1}\right)_{(1/6,1/3)}$
 & $q_i$
 & &
 3 &
$\left(\overline{\boldsymbol{3}},\boldsymbol{1};\boldsymbol{1},\boldsymbol{1}\right)_{(-2/3,-1/3)}$
 & $\bar u_i$
 \\
 3 &
$\left(\boldsymbol{1},\boldsymbol{1};\boldsymbol{1},\boldsymbol{1}\right)_{(1,1)}$
 & $\bar e_i$
 & &
 8 &
$\left(\boldsymbol{1},\boldsymbol{2};\boldsymbol{1},\boldsymbol{1}\right)_{(0,*)}$
 & $m_i$
 \\
 4 &
$\left(\overline{\boldsymbol{3}},\boldsymbol{1};\boldsymbol{1},\boldsymbol{1}\right)_{(1/3,-1/3)}$
 & $\bar d_i$
 & &
 1 &
$\left(\boldsymbol{3},\boldsymbol{1};\boldsymbol{1},\boldsymbol{1}\right)_{(-1/3,1/3)}$
 & $d_i$
 \\
 4 &
$\left(\boldsymbol{1},\boldsymbol{2};\boldsymbol{1},\boldsymbol{1}\right)_{(-1/2,-1)}$
 & $\ell_i$
 & &
 1 &
$\left(\boldsymbol{1},\boldsymbol{2};\boldsymbol{1},\boldsymbol{1}\right)_{(1/2,1)}$
 & $\bar \ell_i$
 \\
 1 &
$\left(\boldsymbol{1},\boldsymbol{2};\boldsymbol{1},\boldsymbol{1}\right)_{(-1/2,0)}$
 & $\phi_i$
 & &
 1 &
$\left(\boldsymbol{1},\boldsymbol{2};\boldsymbol{1},\boldsymbol{1}\right)_{(1/2,0)}$
 & $\bar \phi_i$
 \\
 6 &
$\left(\overline{\boldsymbol{3}},\boldsymbol{1};\boldsymbol{1},\boldsymbol{1}\right)_{(1/3,2/3)}$
 & $\bar\delta_i$
 & &
 6 &
$\left(\boldsymbol{3},\boldsymbol{1};\boldsymbol{1},\boldsymbol{1}\right)_{(-1/3,-2/3)}$
 & $\delta_i$
 \\
 14 &
$\left(\boldsymbol{1},\boldsymbol{1};\boldsymbol{1},\boldsymbol{1}\right)_{(1/2,*)}$
 & $s^+_i$
 & &
 14 &
$\left(\boldsymbol{1},\boldsymbol{1};\boldsymbol{1},\boldsymbol{1}\right)_{(-1/2,*)}$
 & $s^-_i$
 \\
 16 &
$\left(\boldsymbol{1},\boldsymbol{1};\boldsymbol{1},\boldsymbol{1}\right)_{(0,1)}$
 & $\bar n_i$
 & &
 13 &
$\left(\boldsymbol{1},\boldsymbol{1};\boldsymbol{1},\boldsymbol{1}\right)_{(0,-1)}$
 & $n_i$
 \\
 5 &
$\left(\boldsymbol{1},\boldsymbol{1};\boldsymbol{1},\boldsymbol{2}\right)_{(0,1)}$
 & $\bar \eta_i$
 & &
 5 &
$\left(\boldsymbol{1},\boldsymbol{1};\boldsymbol{1},\boldsymbol{2}\right)_{(0,-1)}$
 & $\eta_i$
 \\
 10 &
$\left(\boldsymbol{1},\boldsymbol{1};\boldsymbol{1},\boldsymbol{2}\right)_{(0,0)}$
 & $h_i$
 & &
 2 &
$\left(\boldsymbol{1},\boldsymbol{2};\boldsymbol{1},\boldsymbol{2}\right)_{(0,0)}$
 & $y_i$
 \\
 6 &
$\left(\boldsymbol{1},\boldsymbol{1};\boldsymbol{4},\boldsymbol{1}\right)_{(0,*)}$
 & $f_i$
 & &
 6 &
$\left(\boldsymbol{1},\boldsymbol{1};\overline{\boldsymbol{4}},\boldsymbol{1}\right)_{(0,*)}$
 & $\bar f_i$
 \\
 2 &
$\left(\boldsymbol{1},\boldsymbol{1};\boldsymbol{4},\boldsymbol{1}\right)_{(-1/2,-1)}$
 & $f_i^-$
 & &
 2 &
$\left(\boldsymbol{1},\boldsymbol{1};\overline{\boldsymbol{4}},\boldsymbol{1}\right)_{(1/2,1)}$
 & $\bar f_i^+$
 \\
 4 &
$\left(\boldsymbol{1},\boldsymbol{1};\boldsymbol{1},\boldsymbol{1}\right)_{(0,\pm2)}$
 & $\chi_i$
 & &
 32 &
$\left(\boldsymbol{1},\boldsymbol{1};\boldsymbol{1},\boldsymbol{1}\right)_{(0,0)}$
 & $s^0_i$
 \\
 2 &
$\left(\overline{\boldsymbol{3}},\boldsymbol{1};\boldsymbol{1},\boldsymbol{1}\right)_{(-1/6,2/3)}$
 & $\bar v_i$
 & &
 2 &
$\left(\boldsymbol{3},\boldsymbol{1};\boldsymbol{1},\boldsymbol{1}\right)_{(1/6,-2/3)}$
 & $v_i$
 \\
\hline
\end{tabular}
} \caption{Spectrum. The quantum numbers under $SU(3) \times SU(2)
\times [SU(4) \times SU(2)']$ are shown in boldface; hypercharge and
$B-L$ charge appear as subscripts.  Note that the states $s_i^\pm$,
$f_i$, $\bar f_i$ and $m_i$ have different $B-L$ charges for
different $i$, which we do not explicitly list.} \label{tab:naming3}
\end{table}

When the SM singlet fields (Table \ref{tab:naming3}) obtain VEVs, we
have checked that all vector-like exotic states and unwanted U(1)
gauge bosons obtain mass; leaving only the MSSM states at low
energy.  In addition the $\chi$ fields spontaneously break B-L
leaving over a discrete ${\mathbb Z}_2$ matter parity under which
all quarks and leptons are odd and Higgs doublets are even.  This
symmetry enforces an exact R-parity forbidding the baryon or lepton
number violating operators, $ \bar U \bar D \bar D, \  Q L \bar D, \
L L \bar E, \ L H_u$.

Finally the mu term vanishes in the supersymmetric limit.  This is a
consequence of the fact that the coefficient of the $H_u H_d$ term
in the superpotential has vacuum quantum numbers.  Thus any product
of SM singlets which can appear in the pure singlet superpotential
can appear as an effective mu term.   In fact both the mu term and
the singlet superpotential vanish to order 6 in the product of
fields.   Hence in the supersymmetric vacuum the VEV of the
superpotential and the mu term both vanish. As a consequence, when
supergravity is considered, the supersymmetric vacuum is consistent
with flat Minkowski space. We also obtain non-trivial effective
Yukawa matrices. The charged fermion Yukawa matrices are
\begin{eqnarray}
Y_u&=& \left(
\begin{array}{ccc}
 \widetilde{s}^5 & \widetilde{s}^5 & \widetilde{s}^5 \\
 \widetilde{s}^5 & \widetilde{s}^5 & \widetilde{s}^5 \\
 \widetilde{s}^6 & \widetilde{s}^6 & 1
\end{array}
\right)\;,\quad Y_d~=~ \left(
\begin{array}{ccc}
 \widetilde{s}^5 & \widetilde{s}^5 & 0  \\
 \widetilde{s}^5 & \widetilde{s}^5 & 0  \\
 \widetilde{s}^6 & \widetilde{s}^6 & 0
\end{array}
\right)\;,\quad Y_e~=~ \left(
\begin{array}{ccc}
 \widetilde{s}^5 & \widetilde{s}^5 & \widetilde{s}^6 \\
 \widetilde{s}^5 & \widetilde{s}^5 & \widetilde{s}^6 \\
 \widetilde{s}^6 & \widetilde{s}^6 & 0
\end{array}
\right)\;.\nonumber\\
& &
\end{eqnarray}
where $\tilde s^n$ represents a polynomial in SM singlets beginning
at order $n$ in the product of fields. And we have shown that the
three left-handed neutrinos get small mass due to a non-trivial
See-Saw mechanism involving the 16 right-handed neutrinos and their
13 conjugates. All in all, this ``benchmark" model looks very much
like the MSSM!

Note,   Yukawa couplings,  gauge couplings and  Vector-like exotic
masses are functions  of moduli  (along SUSY flat directions). Some
of these moduli are blow up modes for some,  BUT NOT ALL,  of the
orbifold fixed points.   In fact,  two fixed points are NOT blown
up!

\section{F  theory /  Type IIB}

\begin{figure}
  \includegraphics[height=2.5cm]{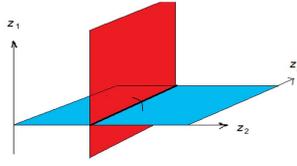}
  \caption{The figure represents 3 complex planes labeled by $z_i, i = 1,2,3$. The four dimensional
  blue surface is the gauge brane and the matter brane is red.  Open strings at the intersection give
  chiral matter in bi-fundamental representations.}
  \label{fig:branes}
\end{figure}

\begin{figure}
  \includegraphics[height=2.5cm]{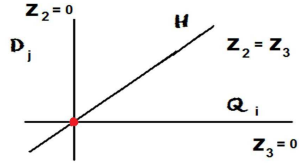}
  \caption{Yukawa couplings are generated at the intersection of two quark branes with a Higgs brane.}
  \label{fig:fermionmasses}
\end{figure}

Now let's change directions and discuss some recent progress in F
theory model
building~\cite{Beasley:2008dc,Beasley:2008kw,Donagi:2008ca,Donagi:2008kj,Blumenhagen:2008zz,Chen:2009me,Marsano:2009ym}.
An $SU(5)$ GUT is obtained on a D7 ``gauge" brane $S \times
R^{3,1}$.  D7 ``matter" branes on $S' \times R^{3,1}$ also exist
with chiral matter in 6D on $\Sigma \times R^{3,1}$ at the
intersection of the gauge and matter branes (Fig. \ref{fig:branes}).
Yukawa couplings enter at the triple intersections $\Sigma_1 \bigcap
\Sigma_2 \bigcap \Sigma_3$ of matter sub-manifolds (Fig.
\ref{fig:fermionmasses}).

$SU(5)$ is broken to the SM gauge group with non-vanishing
hypercharge flux $\langle F_Y \rangle$. Note, this is not possible
in the Heterotic string!  This is because of the term in the
Lagrangian $\int d^{10}x ( dB + A_Y \bigwedge \langle F_Y \rangle
)^2$ which leads to a massive hypercharge gauge boson and
consequently a massive photon. In addition, $\langle F_Y \rangle$ on
the Higgs brane leads to doublet-triplet splitting.  Finally, spinor
representations of $SO(10)$ are possible in F theory; although they
are not possible in the perturbative type IIB string.

It was also demonstrated that a fermion flavor hierarchy is
natural~\cite{Heckman:2008qa,Randall:2009dw}, due to flux in the
$z_2-z_3$ surface breaking geometric flavor symmetry, with Yukawa
matrices of the form
\begin{equation} \lambda \sim  \left(
\begin{array}{ccc} \epsilon^8 & \epsilon^6 & \epsilon^4 \\
\epsilon^6 & \epsilon^4 & \epsilon^2 \\ \epsilon^4 & \epsilon^2 & 1
\end{array} \right)
\end{equation}

Finally, gravity decouples (i.e. $M_{Pl} \rightarrow \infty$) with a
non-compact $z_1$ direction.   These are so-called ``local"
constructions.  A bit of progress has also been made in ``global"
compact constructions~\cite{Blumenhagen:2008zz,Marsano:2009ym}.

\section{Heterotic - F theory duals}

\begin{figure}
  \includegraphics[height=2.5cm]{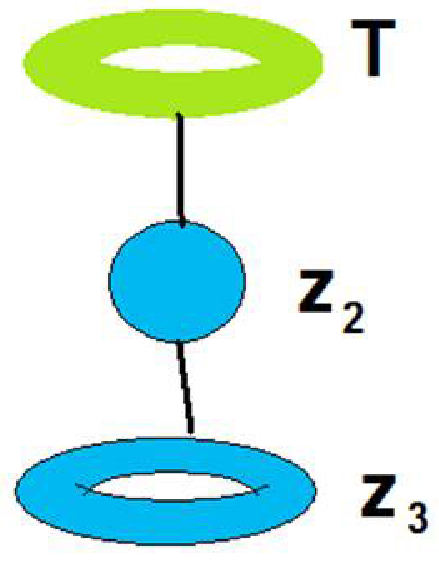}
  \includegraphics[height=2.5cm]{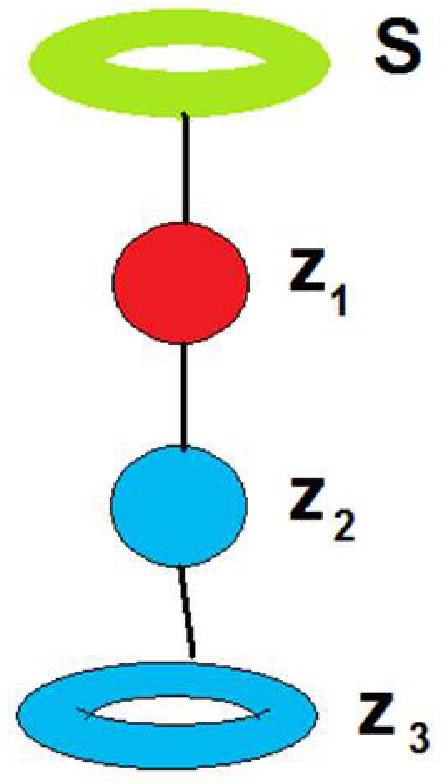}
  \caption{The heterotic string (left) is compactified on the product of two tori fibered over a common 2-sphere (defining a
  Calabi-Yau three-fold), while F theory
  (right) is defined in terms of a torus (whose complex structure defines the coupling strength of a Type IIB string) fibered
  over the product of two 2-spheres and the additional torus fibered over a common 2-sphere (defining a Calabi-Yau four-fold).}
  \label{fig:heteroticvsFtheory}
\end{figure}

F theory defined on a CY$_4$ is dual to the heterotic string defined
on a CY$_3$ (Fig. \ref{fig:heteroticvsFtheory}).  We are now
attempting to construct the F theory dual to our MSSM-like
models~\cite{bobkov}.  The motivation is three-fold.
\begin{enumerate} \item We have also found three family MSSM-like
models using an $SO(5) \times SO(5) \times SO(4)$
torus~\cite{RVW}. This suggests a larger class of MSSM-like
models.  We hope to find a more general description of MSSM-like
models, i.e. all models in the same universality class. \item It
may also provide a general understanding of moduli space, since
from the orbifold view point we must first construct the
superpotential before it is possible to identify the moduli, and
\item It may help us understand moduli stabilization and SUSY
breaking. \end{enumerate}

Let me now discuss why we think the F theory duals may exist.  First
uplift our $E(8) \times E(8)$ orbifold models onto a smooth
Calabi-Yau manifold.   Recall that after the first ${\mathbb Z}_3$
orbifold plus Wilson line $W_3$ we find a 6D $SU(6)$ orbifold GUT
compactified on $(T_2)^2/{\mathbb Z}_3 \times T_2$.  The complete
massless spectrum in this case (including the hidden sector) is
given in Table \ref{tab:6dspectrum}~\cite{Dundee:2008ts}. This
spectrum satisfies the gravity anomaly constraint $N_H - N_V^6 + 29
N_T = 273$, where ($N_H = 320, N_V^6 = 76, N_T = 1$) are the number
of (hyper-, vector,tensor) multiplets.

\begin{table}[h!]
\centering \caption{The full (six dimensional) spectrum of the
``benchmark" model with gauge group $SU(6) \times \left[SO(8) \times
SU(3) \right]'$. Note that $\mathbf{8}_{v+c+s} \equiv \mathbf{8}_{v}
+ \mathbf{8}_{c} + \mathbf{8}_{s}$.   In addition, the states are
written in the language of $D=6,  \ N=1$ supersymmetry.}
\label{tab:6dspectrum} \vspace{5mm}
\begin{footnotesize}
\begin{tabular}{c|c|c}
\hline
Multiplet Type&Representation&Number\\
\hline
\hline
tensor&singlet&1\\
\hline
vector&$(\mathbf{35},1,1)\oplus(1,\mathbf{28},1)$&35 + 28\\
&$\oplus(1,1,\mathbf{8})\oplus5\times(1,1,1)$&8 + 5\\
\hline
hyper&$(\mathbf{20},1,1)\oplus(1,\mathbf{8}_{v+c+s},1)\oplus 4\times(1,1,1)$&20+24+4 \\
&$\oplus9\times\left\{(\mbsix,1,1)\oplus(\mbsixb,1,1)\right\}$&108\\
&$\oplus9\times\left\{(1,1,\mbthree)\oplus(1,1,\mbthreeb)\right\}$&54\\
&$\oplus3\times(1,\mathbf{8}_{v+c+s},1)$&72\\
&$\oplus 36 \times (1,1,1)$&36\\
&SUGRA singlets&2\\
\hline
\end{tabular}
\end{footnotesize}
\end{table}

\begin{figure}
  \includegraphics[height=2.5cm]{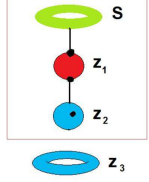}
  \caption{At the 6D level we have a Calabi-Yau three-fold times a torus.  The gauge branes are located at the points $z_1 = 0$ and
  $\infty$.   The matter branes are located schematically at the solid point in $Z_2$.}
  \label{fig:Ftheory2}
\end{figure}

However, using the results of Bershadsky et
al.~\cite{Bershadsky:1996nh} we show that the $E(8) \times E(8)$
heterotic string compactified on a smooth $K_3 \times T_2$, with
instantons imbedded into $K_3$, is equivalent to the orbifolded
theories.   For example, with 12 instantons imbedded into an $SU(3)
\times SU(2)$ subgroup of the first $E(8)$ leaves an $SU(6)$ 6D GUT
with the massless hypermultiplets $(20 + cc) +  18 (6 + c.c.)$. Then
imbedding 12 instantons into an $E(6)$ subgroup of the second $E(8)$
plus additional higgsing leaves an unbroken $SO(8)$ gauge symmetry
with the massless hypermultiplets $4 (8v+ 8s+ 8c + c.c.)$. This is
identical to the massless spectrum of the orbifold GUT, IF we
neglect the additional $SU(3) \times U(1)^5$ symmetry which must be
broken when going to the smooth limit.  In fact, it is expected that
the blow up modes necessary to smooth out the orbifold singularities
will carry charges under some of the orbifold gauge symmetries;
spontaneously breaking these symmetries. Therefore $K_3 \times T_2$
with instantons is the smooth limit of $T_4/{\mathbb Z}_3 \times
T_2$ orbifold plus Wilson line.  In addition, it was shown that F
theory compactified on a Calabi-Yau 3-fold [defined in terms of a
torus $T_2$ fibered over the space $F_n$ $\times T_2$] is dual to an
$E(8) \times E(8)$ heterotic string compactified on $K_3 \times T_2$
with instantons~\cite{Bershadsky:1996nh} (see Fig.
\ref{fig:Ftheory2}).

Pictorially we see that the $SU(6) [SO(8)]$ gauge branes are
localized at the upper [lower] points on the $z_1$ 2-sphere (Fig.
\ref{fig:Ftheory2}).  These 7 branes wrap the four dimensional
surface $S = (z_2,z_3)$.  The matter 7 branes intersect the gauge
branes at points in $z_2$ and wrap the four dimensional surface $S'
= (z_1, z_3)$.   The intersection of the matter and gauge branes is
along the two dimensional surface $\Sigma = (z_3)$.

\begin{figure}
  \includegraphics[height=2.5cm]{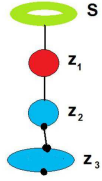}
  \caption{At the 4D level we fiber the last torus over the 2-sphere and retain two orbifold fixed points.  These two
  fixed points are where the two light families are conjectured to be located.  In addition, as long as the fixed points
  remain, the Wilson line wrapping the last torus is stable.}
  \label{fig:Ftheory3}
\end{figure}

We now need to break the 6D $SU(6)$ GUT to $SU(5)$ and then to the
Standard Model.  At the same time we must break the N=1 SUSY in 6D
to N=1 in 4D.   This is accomplished by acting with the ${\mathbb
Z}_2$ orbifold on the torus and the 2-sphere.   A $U(1)$ flux in
$SU(6)$ on the gauge and matter branes breaks $SU(6)$ to $SU(5)$.
The breaking to the Standard Model requires a Wilson line on the
torus. However, we now encounter a possible obstruction to finding
the F theory dual of our Heterotic orbifold model. We need to keep
two orbifold fixed points (Fig. \ref{fig:Ftheory3}) -
\begin{enumerate} \item otherwise hypercharge gets mass~\cite{Nibbelink:2009sp},
\item and the Wilson line shrinks to a point, since $T_2/{\mathbb Z}_2$ is
topologically equivalent to a 2-sphere;
\item  and blow up modes on the heterotic side leave two orbifold fixed
points.
\end{enumerate}

In addition, on the heterotic side the two light families are
located at 4D orbifold fixed points.   We expect that on the F
theory side they will be located on $D_3$ branes fixed at the two
remaining 4D fixed points.

\section{Conclusion} Of course, there are other model building
considerations which I have no time to discuss.  These include,
\begin{itemize}
\item   Gauge coupling unification,
\item Proton decay from dimension 5 and 6 operators,
\item Supersymmetry breaking and sparticle masses (see
\cite{Heckman:2008qa,Aparicio:2008wh}),
\item Moduli stabilization,
\item Cosmological constant and the possible $10^{500}$ vacua; or
\item Cosmology.
\end{itemize}

In conclusion, I believe that it is important to test string theory.
In this talk I have discussed several new ideas in string model
building. I have made the case that orbifold and local GUTs may be
necessary ingredients for finding the MSSM in the string landscape.
Finally, global F theory constructions may open a new window onto
the general MSSM landscape. The first critical test comes with data
at the LHC and the possible discovery of supersymmetry.

\begin{theacknowledgments}
I would like to acknowledge the hospitality and partial support of
the Stanford Institute for Theoretical Physics where this work was
begun. I also received partial support from DOE grant,
DOE/ER/01545-883.
\end{theacknowledgments}

\end{document}